\documentclass[aps,twocolumn,showpacs,prx,amsfonts,amsmath,amssymb,floatfix,groupedaddress,longbibliography]{revtex4-2}
\usepackage{color}
\usepackage{hhline}
\usepackage{mathrsfs}
\usepackage{graphicx}
\usepackage{dcolumn}
\usepackage{bm}
\usepackage{multirow}
\usepackage{booktabs}
\usepackage{afterpage}
\usepackage{amsmath}
\usepackage{amssymb}
\usepackage{mathcomp}

\begin{document}

\title{Illuminating the Bragg intersections as roots of Dirac nodal lines and high-order van Hove singularities}

\author{Ryosuke Akashi$^{1}$}
\thanks{akashi.ryosuke@qst.go.jp}
\affiliation{$^1$Quantum Materials and  Applications Research Center, National Institutes for Quantum Science and  Technology, Ookayama, Meguro-ku, Tokyo, 152-0033, Japan}

\date{\today}
\begin{abstract}
  We theoretically reexamine nearly uniform electron models with weak crystalline potentials. In particular, we theorize the modulation of the plane-wave branches at linear regions where multiple Bragg planes intersect. Any such linear intersections involve three or more plane-wave branches diffracted by the periodic potential. Small inter-branch interactions can yield various crossing and anticrossing singularities with promised breakdown of the quadratic approximation, extending alongside the intersection lines. Most of the intersections run in low-symmetric paths in the Brillouin zone and therefore we cannot completely characterize their electronic states with standard band structure plotting methods. The present theory reveals a general mechanism in nearly uniform systems to induce the Dirac nodal lines and van-Hove singularities with broken quadratic band approximation in three dimensions, which may host a variety of anomalous low-energy electronic properties. We apply the theory to a recently discovered high temperature superconductor H$_{3}$S to interpret the enigmatic density-of-state (DOS) peaking therein. The results show how and {\it why there} the continuous saddle points--the source of the peaked DOS--emerge, as well as reveal the companion Dirac nodal lines hidden in the conduction bands.
\end{abstract}

\maketitle

\section{Introduction}
The crystal structure is the source of various low-energy electronic phenomena, including high-temperature superconductivity. One key effect of that is the formation of the one-particle band structure and resulting nontrivial distribution of the eigenstates in crystal wavenumber and energy space; enhanced density of states, nesting, band flattening, as well as topological structure in eigenfunctions. But relationship between the crystal and band structures is quite nontrivial and, for theory-driven materials design, heuristic approaches are usually utmost effective: Namely, build artificial crystals and execute first principles calculations for those. Useful theories on any such relationships, that may enable design strategies beyond heuristic, have long been desired. 

\begin{figure*}[t!]
  \begin{center}
   \includegraphics[scale=0.30]{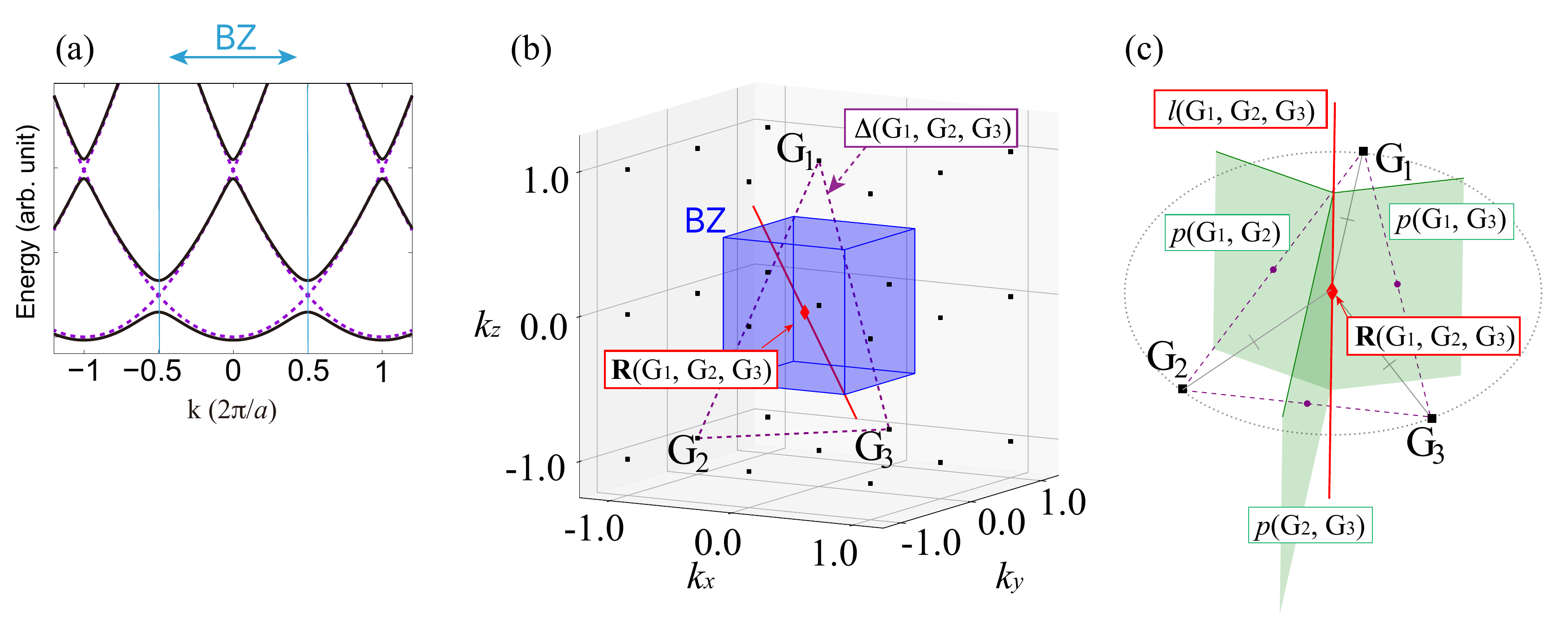}
   \caption{(a) Repeated Brillouin zone scheme and interacting replica plane wave branches in one dimension. (b) Triangle formed by a selected reciprocal points $\Delta({\bf G}_{1}, {\bf G}_{3},{\bf G}_{3})$ and their circumcenter ${\bf R}({\bf G}_{1}, {\bf G}_{3},{\bf G}_{3})$ in three dimensions. (c) Bragg planes $p({\bf G}_{i},{\bf G}_{j})~ (i,j=1,2,3)$ and their intersection $l({\bf G}_{1}, {\bf G}_{3},{\bf G}_{3})$.} 
   \label{fig:parabola-repeated}
   \end{center}
 \end{figure*}

In this study, we seek such a theory by focusing on a geometric aspect of electron as a plane wave under diffraction by periodic potentials. Understanding of the electronic band structure from the diffracted plane waves has been already well established in the Bloch theory~\cite{Bloch1929}, but, in order to find clues toward any missing theories, let us here briefly review the diffraction of the plane waves with elementary concepts. Due to the diffraction potential, the plane waves with (real) wave vector ${\bf k}+{\bf G}$ and ${\bf k}+{\bf G}'$ are hybridized to form eigenstates, where ${\bf G}$ and ${\bf G}'$ denote the reciprocal vector conforming to the system's translational symmetry. When the diffraction is weak, such hybridization occurs only between energetically degenerate states. This mechanism is efficiently visualized by the repeated Brillouin zone scheme~\cite{Ashcroft-Mermin} [Fig.~\ref{fig:parabola-repeated}(a)]. In this view, the wavenumbers at which two diffracted branches meet span Bragg planes. The Bragg planes between the origin and adjacent reciprocal points defines the first Brillouin zone. At the Brillouin zone boundary, there always emerge linearly crossing or anticrossing band pairs. In the latter case, the band gradients have zero components normal to the zone boundary due to symmetry and continuity requirements of the bands. Such anomalies also occur near (not exactly at) the Bragg planes for farther reciprocal points, which may be located in the interior of the Brillouin zone. 

The above description on how the diffracted plane waves interact has been well established, using the nearly free (or uniform, with any mean fields from interactions considered) electron model~\cite{Herman-RMP}. At the early stage of the band theory, this model was useful for developing basic concepts and properties such as Fermi surfaces, as well as served a reference point that demystifies the calculated results of more sophisticated methods like the orthogonalized plane wave method~\cite{Heine-OPW1,Heine-OPW2, Heine-OPW3,Harrison-OPW}. Now the model is mainly referred in textbooks as illustrative examples for the band theory, more than a research object~\cite{Ashcroft-Mermin,Grosso-Parravicini}. We nevertheless point out that how two or more Bragg planes interact has been least understood in the model. The interaction of the multiple plane waves at the Brillouin zone edges has been thoroughly studied already. Extended considerations of the interaction, which can occur in more general regions in the zone, are indeed found to be of revived interest in the modern context, as revealed through the paper. 

By definition, at intersections of the Bragg planes three or more diffracted branches are degenerate and, with weak periodic perturbations, crossing and anticrossing band reconstructions occur there. The interactions between more than two diffracted waves should be far more diverse than on the Bragg planes due to the increased degrees of freedom of relative geometry of the branches and inter-branch matrix elements. Thus the intersections of the Bragg planes are expected to induce various kinds of linearly extended crossing or anticrossing band critical points in the Brillouin zone. Notably, general properties that are only determined by the crystal geometry should be found in the limit of weak diffractions. The colalescence of plane wave branches at the hexagonal K edges has been considered~\cite{Harrison_zinc_PR1962,Bennett_Falicov_zinc_PR1964} in relation to anomalous quantum oscillations observed experimentally~\cite{Dhillon_Shoenberg_zinc_deHaas}. The present theory constitutes a basis for generalization of this work.

Motivated by the above expectations, in this paper we explore the electronic theory at the intersections of the Bragg planes, that applies to any periodic crystals. As an illustrative example, we derive the positions of the intersection lines for the cubic systems, many of which are found to run along low-symmetry paths away from standard ${\bf k}$-point paths for the band structure calculations. We show that the Bragg intersections can host the Dirac nodal lines~\cite{Mullen_DNL_PRL2015,Kim_DNL_PRL2015,Fang_DNL_PRB2015}, as well as linearly extended band extrema in their vicinities. In particular, we indeed find an example where the Bragg intersections may be the origin of linearly extended saddle critical points and enhanced van Hove singularity in the density of states~\cite{vHS}: cubic H$_{3}$S~\cite{Eremets}, known as 200~K superconductor. The present theoretical considerations encourage studies on electronic states under weak periodic potentials with a renewed interest, as well as an overhaul of band structure analysis methods, both of which will be useful for theoretical materials design.






\section{Theory}
Hereafter we consider the diffraction of the plane waves with the repeated zone scheme mentioned above. There, the branches of the diffracted waves are conveniently represented as the band replicas shifted by reciprocal vectors ${\bf G}$; or as a shorthand notation each ${\bf G}$ represents a diffracted planewave branch. Usually the Bragg planes are defined as the bisection of the line between the origin and a reciprocal point ${\bf G}$. For convenience we adopt a duplicate definition of the Bragg plane as a bisector of the line connecting two reciprocal points ${\bf G}_{1}$ and ${\bf G}_{2}$, by
\begin{eqnarray}
  {\bf k}\in p({\bf G}_{1},{\bf G}_{2}) \Leftrightarrow | {\bf k}-{\bf G}_{1}|=| {\bf k}-{\bf G}_{2}
  |.
   \end{eqnarray}


Let us substantiate the concept of intersection of Bragg planes. Since the Bragg planes are bisections between two reciprocal points, their linear intersection is equidistant from three or more reciprocal points. Conversely, every triad of reciprocal points $\{ {\bf G}_1, {\bf G}_2, {\bf G}_3\}$ that forms a triangle specifies its corresponding intersection. Take three reciprocal vectors ${\bf G}_{1}$, ${\bf G}_{2}$ and ${\bf G}_{3}$ and define the triangle formed by them as $\Delta({\bf G}_{1}, {\bf G}_{2}, {\bf G}_{3})$. The Bragg planes $p({\bf G}_1, {\bf G}_2)$, $p({\bf G}_2, {\bf G}_3)$ and $p({\bf G}_3, {\bf G}_1)$ cross at a single common line. The intersection line, termed $l({\bf G}_{1},{\bf G}_{2},{\bf G}_{3})$, is normal to $\Delta({\bf G}_{1}, {\bf G}_{2}, {\bf G}_{3})$ and passes through its circumcenter ${\bf R}({\bf G}_{1}, {\bf G}_{2}, {\bf G}_{3})$. Since the energy of branches are determined by the radius measured from the corresponding reciprocal points, three or more diffracted replica branches exactly cross on the intersection lines. The geometric relations of the objects defined here are illustrated in Fig.~\ref{fig:parabola-repeated} (b)(c).


\begin{figure}[t!]
  \begin{center}
   \includegraphics[scale=0.32]{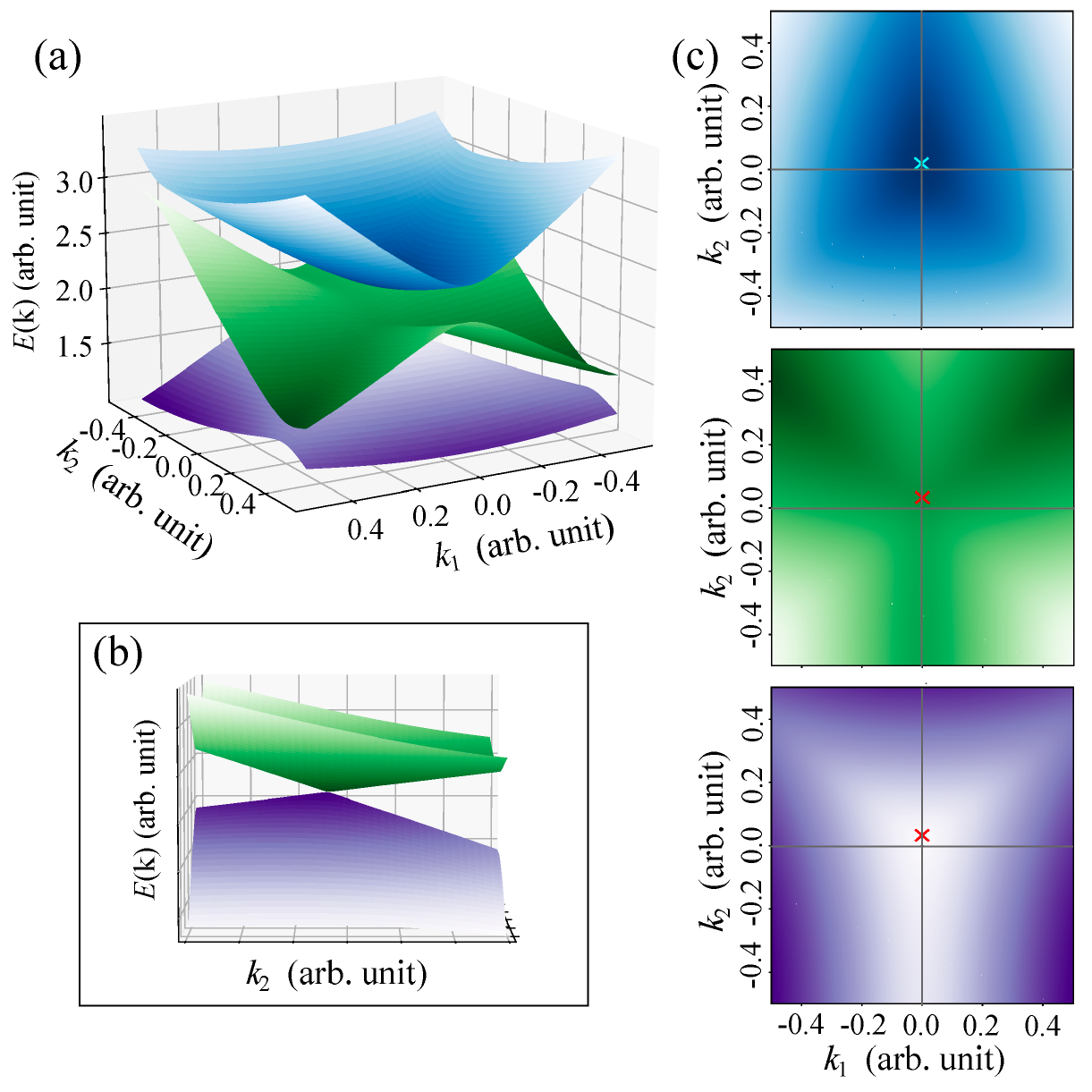}
   \caption{Typical behavior of three energy surfaces emerging at the intersection. Model parameters were set to ${\bf G}_{1}=(2, 0),  {\bf G}_{2}=(2\cos 105^{\circ}, 2\sin 105^{\circ}), {\bf G}_{3}=(2\cos 255^{\circ}, 2\sin 255^{\circ})$ and $V_{12}=V_{31}=0.1, V_{23}=0.15$, respectively.  (b) Close up view of the lower two bands near the Dirac point. (c)--(e) two-dimensional heat map of the bands, where crosses indicate the critical and Dirac points, respectively.} 
   \label{fig:three-bands-surface}
   \end{center}
 \end{figure}

Let us next examine the two-dimensional bandstructure on $\Delta({\bf G}_{1}, {\bf G}_{2},{\bf G}_{3})$ near the intersection $l({\bf G}_{1},{\bf G}_{2},{\bf G}_{3})$ with introduction of weak diffraction potential. The band structure is modeled by the three-state Hamiltonian~\footnote{We do not address the case of two plane-wave interaction at the Bragg planes $p({\bf G}_{
1},{\bf G}_{2})$ as it is trivial: There form linearly crossing bands at the whole planes if the corresponding Fourier component with wavenumber ${\bf G}_{1}-{\bf G}_{2}$ of the potential is zero, and form anticrossing otherwise~\cite{Ashcroft-Mermin}.}
\begin{eqnarray}
  H({\bf k})
  =
  \left(
  \begin{array}{ccc}
  E_{0}({\bf k}-{\bf G}_{1}) & V_{12} & V_{31}^{\ast} \\
  V_{12}^{\ast} & E_{0}({\bf k}-{\bf G}_{2}) & V_{23} \\
  V_{31} & V_{23}^{\ast} & E_{0}({\bf k}-{\bf G}_{3})
  \end{array}
  \right)
  \label{eq:threefold-Hamil}
\end{eqnarray}
with $E_{0}({\bf k})={\bf k}^{2}/2$. The energy eigenvalue of $H({\bf k})$ is referred to as $E({\bf k})$ later. Throughout this section we assume the time-reversal and periodic translational symmetries but no other spatial symmetries unless declared. This model can generate various structures; here we show some typical ones.

With weak nondiagonal components, three crossing bands evolves into partially anticrossing bands. As long as all the angles of $\Delta({\bf G}_{1}, {\bf G}_{2},{\bf G}_{3})$ are smaller than $90^{\circ}$, their in-plane critical points, where the band gradient $\partial E({\bf k})/\partial {\bf k}$ is zero in the direction of $\Delta({\bf G}_{1}, {\bf G}_{2},{\bf G}_{3})$, are located near ${\bf R}({\bf G}_{1},{\bf G}_{2},{\bf G}_{3})$. For certain but broad range of the nondiagonal values (we revisit this point later), a degeneracy of Dirac type can also occur near $l({\bf G}_{1},{\bf G}_{2},{\bf G}_{3})$. This Dirac point can be either in upper or lower two bands of the three, depending on the signs of the nondiagonal components. In the vicinity of those anomalies, the quadratic expansion of $E({\bf k})-E({\bf R}({\bf G}_{1}, {\bf G}_{2},{\bf G}_{3}))$ is obviously broken down [Fig.~\ref{fig:three-bands-surface}].

The current analysis trivially applies to any planes normal to $l({\bf G}_{1},{\bf G}_{2},{\bf G}_{3})$. With a reasonable smooth k-dependence of the nondiagonal components, the anomalies, Dirac or non-quadratic in-plane critical points, therefore extend continuously along $l({\bf G}_{1},{\bf G}_{2},{\bf G}_{3})$, which is the key concept of the current theory. 

The expected continuous in-plane critical points generally have nonzero dispersions in the direction of $l({\bf G}_{1},{\bf G}_{2},{\bf G}_{3})$. If they exhibit extrema along that, nevertheless, they are almost always anomalous critical points where the rank of the inverse mass matrix is smaller than three. Such critical points induce the van Hove singularities in the density of states~\cite{vHS} with orders of singularitie stronger than expected in the spatial dimensions of ${\bf k}$~(Ref.~\onlinecite{Yuan-Fu-vHS}), termed higher-order van Hove singularities. We always expect such critical points near ${\bf R}({\bf G}_{1}, {\bf G}_{2},{\bf G}_{3})$ since all the noninteracting replica branches $E_{0}({\bf k}-{\bf G}_{i})$ $(i=1,2,3)$ are trivially convex in the $l({\bf G}_{1},{\bf G}_{2},{\bf G}_{3})$ direction. If the dispersion happens to be small, the singularity should be more prominent.


We also note about intersections of the intersection lines. Any intersection lines cross anywhere with others related by some symmetry (e.g., at the BZ surface). The replica branches involved in each intersection lines all interact at such crossing points, by which the continuous anomalies of the respective intersections are connected, smoothly unless additional degeneracies hold. As a result, looped anomalies running across the intersection lines emerge.

To sum up, any three replica plane waves are degenerate at the threefold Bragg intersections, at which they interact with each other to form continuous band anomalies running at its proximity. Remarkably, most of the intersections are not located along the high-symmetry lines. The automated band calculation procedure using the standardized ${\bf k}$-path definitions cannot reveal the whole shape of the continuous anomalies related to the intersections since the calculations run only along selected high symmetry lines. To capture the true figure of the anomalies, nonstandard band analysis methods are needed as we demonstrate below.

\subsection*{Dirac Nodal lines}
Considering the recent intensive attention to the Dirac-type degeneracies, we here discuss how the Bragg intersections yield Dirac nodal lines. Asano and Hotta proved a condition with which the Dirac degeneracy is ``feasible'', or persistent against continuous modification to the Hamiltonian~\cite{Asano_Hotta_PRB2011}. Namely, the $n_{\rm d}$-dimensional $m$-fold Dirac-type band contact in $n_{\rm u}$ dimensional parameter space is feasible if
\begin{eqnarray}
  n_{\rm d}=n_{\rm u}-m^2+1+n_{\rm c} \geq 0
  ,
  \label{eq:Asano-Hotta}
\end{eqnarray}
with $n_{\rm c}$ being the number of constraints among the components of the Hamiltonian imposed by symmetries of the system.

Let us take the three-state Hamiltonian of the form Eq.(\ref{eq:threefold-Hamil}) under time reversal symmetry. The constraints among the components are governed by the ${\bf k}$ group of the respective ${\bf k}$ points~\cite{Ouyou-gunron-Eng}. When the system has inversion symmetry, all the ${\bf k}$ points are invariant under successive spatial inversion and time reversal operations. One constraint is then imposed $(n_{\rm c}=1)$~\cite{Herring_degeneracy_PR1937,Asano_Hotta_PRB2011}. Therefore, in three dimensions the two-fold Dirac degeneracy lines $(n_{\rm u}=3, m=2, n_{\rm d}=1)$ are, if present, feasible. This fact is also interpretable as the protection by the Berry phase being discrete~\cite{Hirayama2017}. 

The presence of the Dirac nodal lines along any specific intersection $l({\bf G}_{1}, {\bf G}_{2},{\bf G}_{3})$ crucially depends on the periodic potential and choice of ${\bf G}_{1}, {\bf G}_{2}, {\bf G}_{3}$, but we can at least infer below that the nodal lines may emerge with considerable probability if the systems is inversion symmetric. For this, we consider the Hamiltonian
\begin{eqnarray}
  H=
  \begin{pmatrix}
    E_{0}({\bf k}-{\bf G}_{1}) &\ a &\ a \\
    a &\ E_{0}({\bf k}-{\bf G}_{2}) &\ a \\
    a &\ a &\ E_{0}({\bf k}-{\bf G}_{3})
  \end{pmatrix},
  \label{eq:3sym-Hamiltonian}
\end{eqnarray}
with $a$ being an arbitrary real parameter. The Hamiltonian Eq.(\ref{eq:threefold-Hamil}) has this constrained form when the system is symmetric under threefold rotation around $l({\bf G}_{1}, {\bf G}_{2},{\bf G}_{3})$ and the reciprocal points $\{{\bf G}_{1}, {\bf G}_{2},{\bf G}_{3}\}$ are permuted by that rotation. On $l({\bf G}_{1}, {\bf G}_{2},{\bf G}_{3})$, $E_{0}({\bf k}-{\bf G}_{1})=E_{0}({\bf k}-{\bf G}_{2})=E_{0}({\bf k}-{\bf G}_{3})\equiv E$ and the Hamiltonian has two-fold degenerate eigenstates with eigenvalue $E-a$ and one nondegenerate eigenstate with eigenvalue $E+2a$. The former solution is found to be the Dirac nodal line. Suppose next we modify the crystal gradually. Even if the rotational symmetry is broken then, the specific values of the Hamiltonian components will vary only continuously. With this premise and the guaranteed feasibility of the Dirac nodal line, we get to a conclusion: Any three-state Hamiltonian specified by points ${\bf G}_{1},{\bf G}_{2},{\bf G}_{3}$ Eq.(\ref{eq:threefold-Hamil}) and consistent with the spatial inversion symmetry generates the Dirac nodal lines that originate from Eq.~(\ref{eq:3sym-Hamiltonian}), unless they experience interactions with other nodal lines like merging through the continuous modification of the Hamiltonian. The actual values of the Hamiltonian components are of course affected by the plane-wave states other than the three via perturbations, but still similar arguments hold: The perturbative corrections due to the other modes can in principle be analyzed using the Brillouin-Wigner treatment~\cite{Asano_Hotta_PRB2011}. 

\begin{table*}[b!]
  \caption[t]
  {List of intersection lines in cubic lattices, determined by three reciprocal vectors ${\bf G}_{1}, {\bf G}_{2}, {\bf G}_{3}$. Lines are sorted in ascending order up to 15th by the (squared) circumladius, which is proportional to the energy minimum along the intersection. Unit in the reciprocal space is taken to be $2\pi/a$, with $a$ being the cubic lattice parameter. Positions are in Cartesian coordinate. The list does not include those with $\Delta({\bf G}_{1}, {\bf G}_{2}, {\bf G}_{3})$ being obtuse triangle, for which the critical point should not occur near $l({\bf G}_{1}, {\bf G}_{2}, {\bf G}_{3})$. Symmetry C$_{1v}$ is equivalent to C$_{s}$ in the standard notation.}
  \begin{center}
  \label{tab:paths}
  \tabcolsep = 1mm
  \begin{tabular}{|l |c|c|c|c|c|c|c|c|c|} \hline
  Label & (Radius)$^2$  & ${\bf G}_{2}-{\bf G}_{1}$ & ${\bf G}_{3}-{\bf G}_{1}$& Center&Direction &Conventional path &Symmetry& BCC & FCC \\ \hline 
  $l(1)$ & 1/2 & (1, 1, 0) & (1, 0, 0) &(1/2, 1/2, 1)&(0, 0, 1)& \checkmark &C$_{4v}$ &            &  \\ \hline
  $l(2)$ & 2/3 & (1, 1, 0) & (1, 0, 1) &(-1/3, 1/3, 1/3)&(-1,1,1)& \checkmark &C$_{3v}$& \checkmark &  \\ \hline
  $l(3)$ & 3/4 & (1, 1, 1) & (1, 1, 0) &(1/2, 1/2, 1/2)&(1, -1,0)& \checkmark &C$_{2v}$&            &  \\ \hline
  $l(4)$ & 1 & (1, 1, 0) & (1, -1, 0) &(0,0,0)&(0,0,1)& \checkmark  &C$_{4v}$& \checkmark &  \\ \hline
  $l(5)$ & 9/8 & (1, 1, 1) & (1, 1, -1) &(-1/4, -1/4, 0)&(1, -1, 0)&           &C$_{1v}$&            & \checkmark \\ \hline
  $l(6)$ & 5/4 & (1, 1, 1) & (1, 0, 2) &(1/2, 0, 0)&(-2, 1, 1)&            &C$_{1v}$&            &  \\ \hline
  $l(7)$ & 25/18 & (1, 1, 0) & (1, 0, 2) &(-5/18, 5/18,-2/18)&(-2, 2, 1)&          &C$_{1v}$&            &  \\ \hline
  $l(7')$ & 25/18 & (1, 0, 2) & (-1, 0, 1) &(1/6, 0, 1/6)&(0, 1, 0)&          &C$_{1v}$&            &  \\ \hline
  $l(8)$ & 3/2 & (1, 1, 2) & (1, 1, 0) &(1/2, 1/2, 0)&(1, -1, 0)& \checkmark &C$_{2v}$& \checkmark &  \\ \hline
  $l(8')$ & 3/2 & (1, 1, 2) & (1, 0, 2) &(1/2, 1/2, 0)&(-2, 0, 1)& \checkmark &C$_{1v}$& \checkmark &  \\ \hline
  $l(9)$ & 25/16 & (1, 0, 2) & (-1, 0, 2) &(0, 0, 1/4)&(0, 1, 0)&         &C$_{1v}$&            &  \\ \hline
  $l(10)$ & 45/28 & (1, 1, 1) & (0, -1, 2) &(5/14, -1/14, 3/14)&(-3, 2, 1)&        &C$_{1}$&            &  \\ \hline
  $l(11)$ & 18/11 & (1, 1, 2) & (0, -1, 1) &(5/11, 2/11, 2/11)&(-3,1,1)&        &C$_{1v}$& \checkmark &  \\ \hline
  $l(12)$ & 25/14 & (1, 1, 2) & (0, -1, 2) &(5/14,1/14, -4/14)&(-4, 2, 1)&        &C$_{1}$&            &  \\ \hline
  $l(13)$ & 9/5 & (1, 1, 2) & (1, -1, 2) &(-2/5, 0, 1/5)&(-2, 0, 1)&          &C$_{1v}$& \checkmark &  \\ \hline
  $l(14)$ & 2 & (1, 1, 2) & (1, 0, 1) &(0, 0, 0)&(-1, -1, 1)& \checkmark  &C$_{3v}$& \checkmark &  \\ \hline
  $l(15)$ & 9/4 & (1, 1, 2) & (1, 1, 1) &(0, 0, 1/2)&(-1, 1, 0)& \checkmark & C$_{2v}$&           &  \\ \hline
  \end{tabular}
  \end{center}
  \end{table*}

Symmetries other than the inversion can serve to confine the position of the nodal line to certain regions. For example, suppose a system's mirror symmetry leaves a point ${\bf G}_{1}$ invariant and interchanges ${\bf G}_{2}$ and ${\bf G}_{3}$. The ${\bf k}$ group at all points on $p({\bf G}_{2}, {\bf G}_{3})$ then includes that mirror. The Dirac nodal line that originates from the Hamiltonian Eq.~(\ref{eq:3sym-Hamiltonian}) remains to be on this plane. This is easily proved. The allowed form of the Hamiltonian on this plane is expressed as follows with additional real parameters $x$ and $y$
\begin{eqnarray}
  &&H=
  \nonumber \\
  &&\begin{pmatrix}
    E_{0}({\bf k}-{\bf G}_{1}) &\ a &\ a \\
    a &\ E_{0}({\bf k}-{\bf G}_{1})+x({\bf k}) &\ a+y \\
    a &\ a+y &\ E_{0}({\bf k}-{\bf G}_{1})+x({\bf k})
  \end{pmatrix}
  ,
  \nonumber \\
\end{eqnarray}
where we have noticed that $E_{0}({\bf k}-{\bf G}_{2})=E_{0}({\bf k}-{\bf G}_{3})\equiv E_{0}({\bf k}-{\bf G}_{1})+x({\bf k})$ by the symmetry. For given $y$, twofold degeneracy occurs when $x({\bf k})=y(y+2a)/(y+a)$, which can be realized by moving the $k$ point in either direction from $l({\bf G}_{1},{\bf G}_{2},{\bf G}_{3})$ along $p({\bf G}_{2}, {\bf G}_{3})$. Continuation of this degeneracy forms the Dirac nodal line. Unless the original nodal line experience the merging, this nodal line is the one rooted from the former. The situation is also understood by the Asano-Hotta theorem Eq.(\ref{eq:Asano-Hotta}) with $n_{u}=2$ and $n_{c}=2$.

\section{Examples}
To reveal the actual behavior of the intersections in crystals, we analyze the nearly uniform electron model~\cite{Ashcroft-Mermin}. The model is described by Hamiltonian with the plane-wave basis set as
\begin{eqnarray}
  && H({\bf k}) = 
  \nonumber \\
  && \left(
    \begin{array}{ccccc}
    \ddots \hspace*{5pt} &&&& \\
    \hspace*{5pt}&E_{0}({\bf k}-{\bf G}_{1}) \hspace*{5pt}  &  V({\bf G}_{1}\!-\!{\bf G}_{2}) \hspace*{5pt}  & V({\bf G}_{1}\!-\!{\bf G}_{3}) \hspace*{5pt}& \\
    \hspace*{5pt}&V^{\ast}({\bf G}_{1}\!-\!{\bf G}_{2}) \hspace*{5pt} & E_{0}({\bf k}-{\bf G}_{2}) \hspace*{5pt} & V({\bf G}_{2}\!-\!{\bf G}_{3}) \hspace*{5pt}& \\
    \hspace*{5pt}&V^{\ast}({\bf G}_{1}\!-\!{\bf G}_{3}) \hspace*{5pt}& V^{\ast}({\bf G}_{2}\!-\!{\bf G}_{3}) \hspace*{5pt} & E_{0}({\bf k}-{\bf G}_{3}) \hspace*{5pt}& \\
    &&&& \ddots \hspace*{5pt} \\
    \end{array}
  \right)
  \nonumber \\
  \label{eq:nearly-uniform}
\end{eqnarray}
where $\{ {\bf G} \}$ conforms to the corresponding translational symmetry. 

The crystalline potential, having the translational symmetry as well, is characterized by the Fourier components
\begin{eqnarray}
  V({\bf r})=\sum_{{\bf G}} e^{i {\bf G}\cdot {\bf r}}V({\bf G})
  .
\end{eqnarray}
In the uniform limit $[V({\bf r})\rightarrow 0]$, the band structure corresponds to that of the free electron folded into the Brillouin zone (empty lattice model). 

\begin{figure}[h]
  \begin{center}
   \includegraphics[scale=0.25]{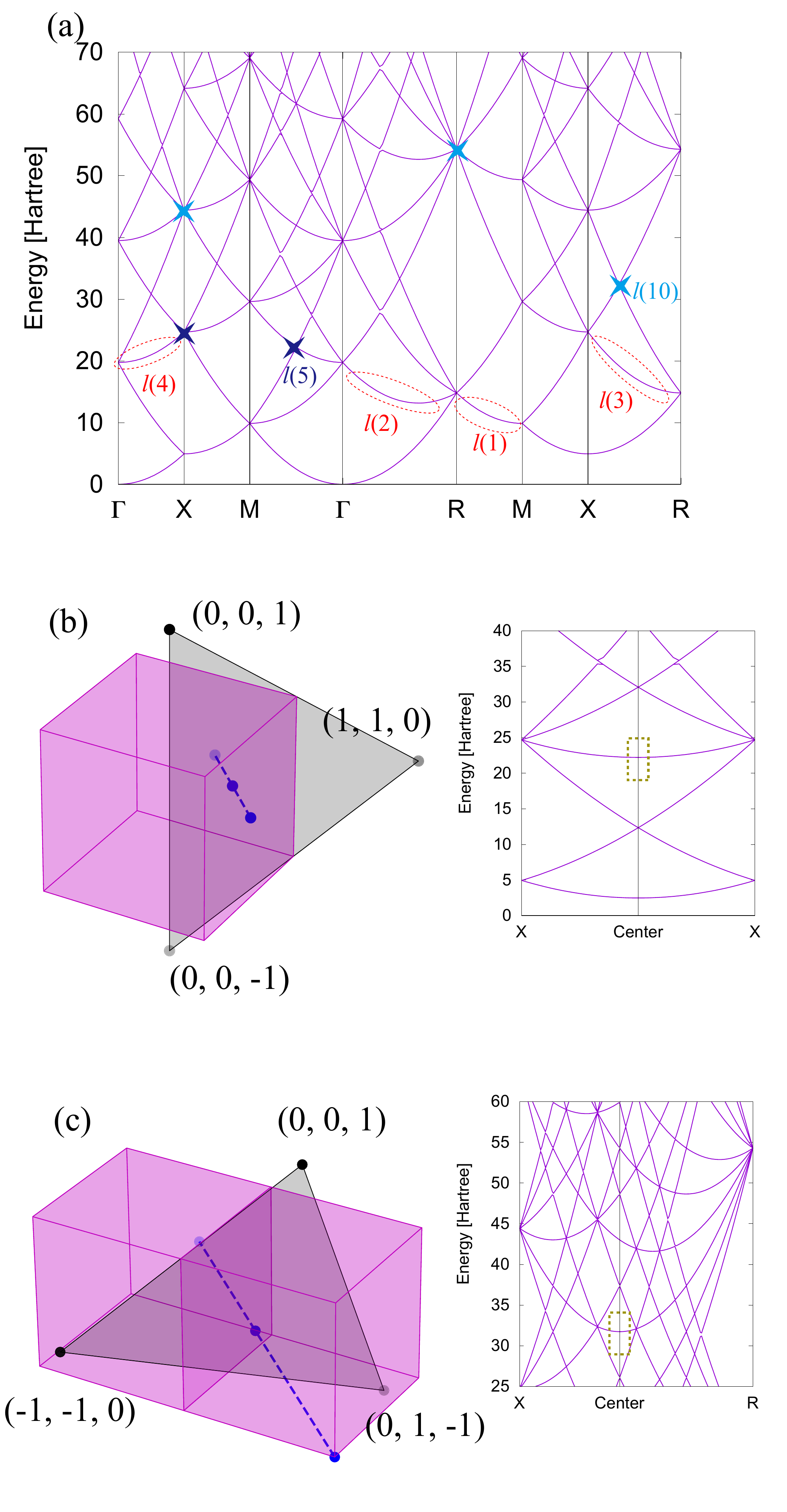}
   \caption{(a) The band structure of empty lattice model on the simple cubic lattice. Representative intersection lines $l(1)$--$l(4)$ and cross sections of $l(5)$ and $l(10)$ are marked. The cubic lattice parameter $a=1.0$ Bohr. (b)(c) Positions of $l(5)$ and $l(10)$ in the Brillouin zone, band structure along those without lattice potential. Dashed boxes indicate the threefold degeneracy and planar regions on which the interacting band structures are displayed in Fig.~\ref{fig:simple-cube-band-surface}} 
   \label{fig:simple-cube-band}
   \end{center}
 \end{figure}

Hereafter all the examples concerned are cubic but let us remember that our general theory is obviously applicable to any crystalline systems. The discussions other than the Dirac nodal lines hold even without the inversion symmetry.

\subsection{Cubic lattices}
For example, we here derive the intersection lines in a simple cubic Bravais lattice. We performed an exhaustive search of triads of reciprocal lattice points that form triangles. The triads are listed in Table~\ref{tab:paths} with their circumcenter positions and directions of the intersection lines.


  \begin{figure}[h!]
    \begin{center}
     \includegraphics[scale=0.35]{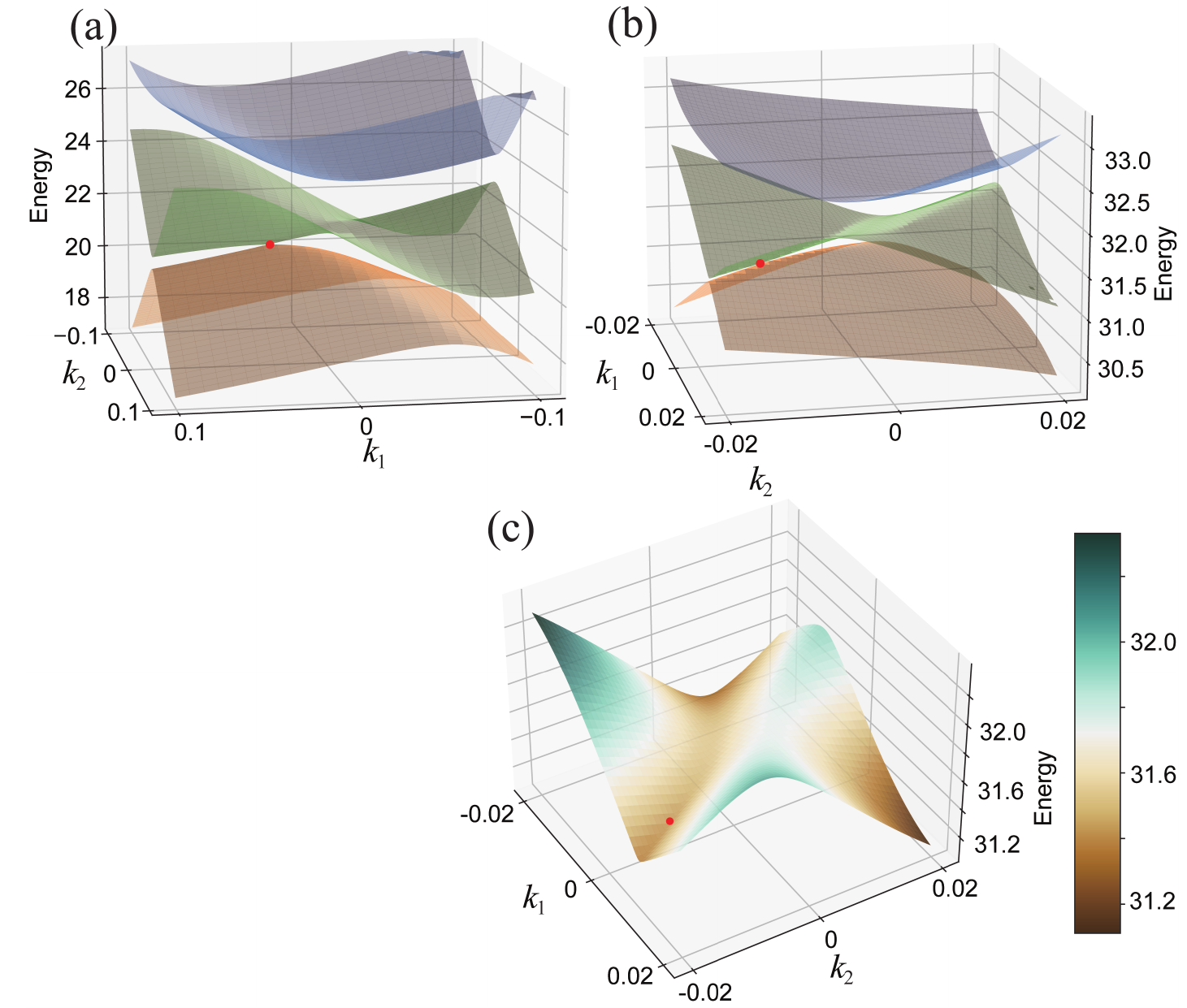}
     \caption{Band structures on a surface normal to the intersection lines (a) $l(5)$ and (b) $l(10)$. The Dirac point is indicated by circle. The interactions are set for (a) $V_{12}=V_{31}=V(|{\bf G}|=\sqrt{3})=1.0$, $V_{23}=V(|{\bf G}|=2)=0.5$ and (b) $V(|{\bf G}|=\sqrt{3})=0.4$, $V(|{\bf G}|=\sqrt{5})=0.2$, $V(|{\bf G}|=\sqrt{6})=0.1$, respectively. (c) Close-up view of the second band in panel (b), with its eigenvalues as a heatmap.} 
     \label{fig:simple-cube-band-surface}
     \end{center}
   \end{figure}

Seen by ascending order with respect to the squared radius, the first four lines $l(1)$--$l(4)$ are along the special high symmetry paths in the Brillouin zones~(see reviews, e.g., Ref.~\cite{Curtarolo} for path labeling), which can be found by the standard band calculations. The first nontrivial intersection is $l(5)$, whose position is depicted in Fig.~\ref{fig:simple-cube-band}(b). We also indicate in Table if each intersection is on the standard k-point paths. 

We plot the free dispersion $E_{0}({\bf k})={\bf k}^2/2$ folded in the Brillouin zone (empty lattice model) [Fig.\ref{fig:simple-cube-band}(a)]. The intersection appears there as threefold degenerate bands. Some intersections are on the standard k-point paths: we mark repersentatives of such lines $l(1)$--$l(4)$ in the plot. On the other hand, other intersections are not running along the standard paths and we can see only their cuts as points, which we also mark for representative lines $l(5)$ and $l(10)$. Calculations on non-standard paths are necessary for revealing the whole structure of the threefold degenerate bands, as shown in Fig.~\ref{fig:simple-cube-band}(b)(c).

The degeneracy of the threefold bands along the intersection is generally lifted by infinitesimal crystalline potential in diverse ways. We are not doing thorough examinations of that but instead seeing two representative cases, $l(5)$ and $l(10)$. 

The intersection $l(5)$ is joined by three branches originating from ${\bf G}_{1}=(1, 1, 0)$, ${\bf G}_{2}=(0, 0, 1)$, ${\bf G}_{3}=(0, 0, -1)$, which form isosceles triangle. An example set of interactions (see caption in Fig.~\ref{fig:simple-cube-band-surface}) modifies the crossing branches to form several in-plane critical points. The branches are shown in upper panel of Fig.~\ref{fig:simple-cube-band-surface}(a). Because of the spatial inversion and mirror symmetry by $p({\bf G}_{2}, {\bf G}_{3})$, a Dirac point is then ensured {\it somewhere} near the intersection on $p({\bf G}_{2}, {\bf G}_{3})$. For other planes parallel to $\Delta({\bf G}_{1},{\bf G}_{2}, {\bf G}_{3})$, the in-plane Hamiltonian is essentially the same with slight changes to the diagonal components $E({\bf k}-{\bf G}_{i})$. The existence of the Dirac and in-plane critical points near those on $\Delta({\bf G}_{1},{\bf G}_{2}, {\bf G}_{3})$ is ensured. Therefore, there forms a dispersive Dirac nodal line in the vicinity of $l(5)$ as well as continuous lines with zero in-plane gradients. 

The intersection $l(10)$ is joined by branches ${\bf G}_{1}=(0, 0, 1)$, ${\bf G}_{2}=(-1, -1, 0)$, ${\bf G}_{3}=(0, 1, -1)$ forming a scalene triangle. This line runs straight across X and R points on different Brillouin zone replicas in a fractional direction [Fig.~\ref{fig:simple-cube-band}(c)]. We plot the band structure on $\Delta({\bf G}_{1},{\bf G}_{2}, {\bf G}_{3})$ in Fig.~\ref{fig:simple-cube-band-surface}~(b). Although there is no obvious symmetry among the inter-branch interactions, we found a Dirac point near $l({\bf G}_{1},{\bf G}_{2}, {\bf G}_{3})$ as indicated. Considering the inversion symmetry of the system, this probably originates from that in the limit $V_{12}=V_{23}=V_{31}$ as discussed in the previous section. Because of the low symmetry of $\Delta({\bf G}_{1},{\bf G}_{2}, {\bf G}_{3})$, the in-plane critical points appear in a complicated way. We do not explore any general remarks on how they form but just do show a reprensetative band with several in-plane extrema and saddles and a Dirac point in Fig.~\ref{fig:simple-cube-band-surface}~(c), that must extend along $l(10)$. Note that the departure from the quadratic shape is clearly observed as expected.

The significance of these anomalies depends also on the band dispersion along the intersections, which should also be a common property determined by only the spatial symmetry. In the above case $l(5)$ show smaller dispersions compared with $l(10)$ [See Fig.~\ref{fig:simple-cube-band}]. In the next subsection, we see an example material where continuous saddles emerging from an intersection with small dispersion yield a higher-order van Hove singularity appearing as a sharp peak in the density of states.




\subsection{Cubic H$_{3}$S at high pressure}

\begin{figure}[t!]
 \begin{center}
  \includegraphics[scale=0.45]{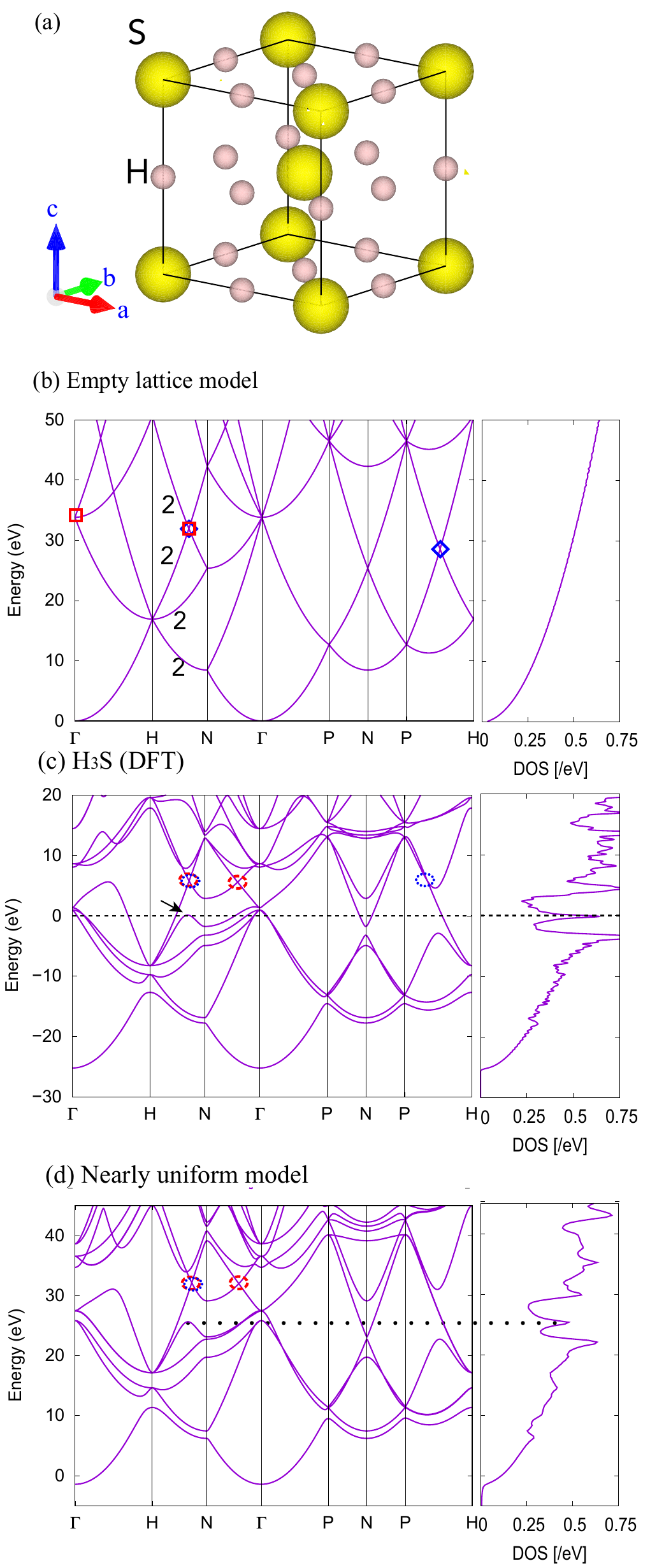}
  \caption{(a) Crystal structure of BCC-H$_{3}$S. (a) Band structure in empty lattice model in BCC lattice ($a=5.6367$ Bohr). Cuts of $l(13)$ and $l(11)$ (see text) are indicated by square and diamond, respectively. Numbers are degrees of degeneracy of the bands. (b) First-principles band structure in BCC-H$_3$S, with the interested extremum indicated by arrow. Ovals highlight the cuts of the Dirac nodal lines. (c) That with the nearly uniform model with selected interaction parameters. Dotted line is a guide to the eye for correspondence of the extremum and DOS peak.} 
  \label{fig:bands_NF_BCC}
  \end{center}
\end{figure}


 \begin{figure*}[t!]
  \begin{center}
   \includegraphics[scale=1.0]{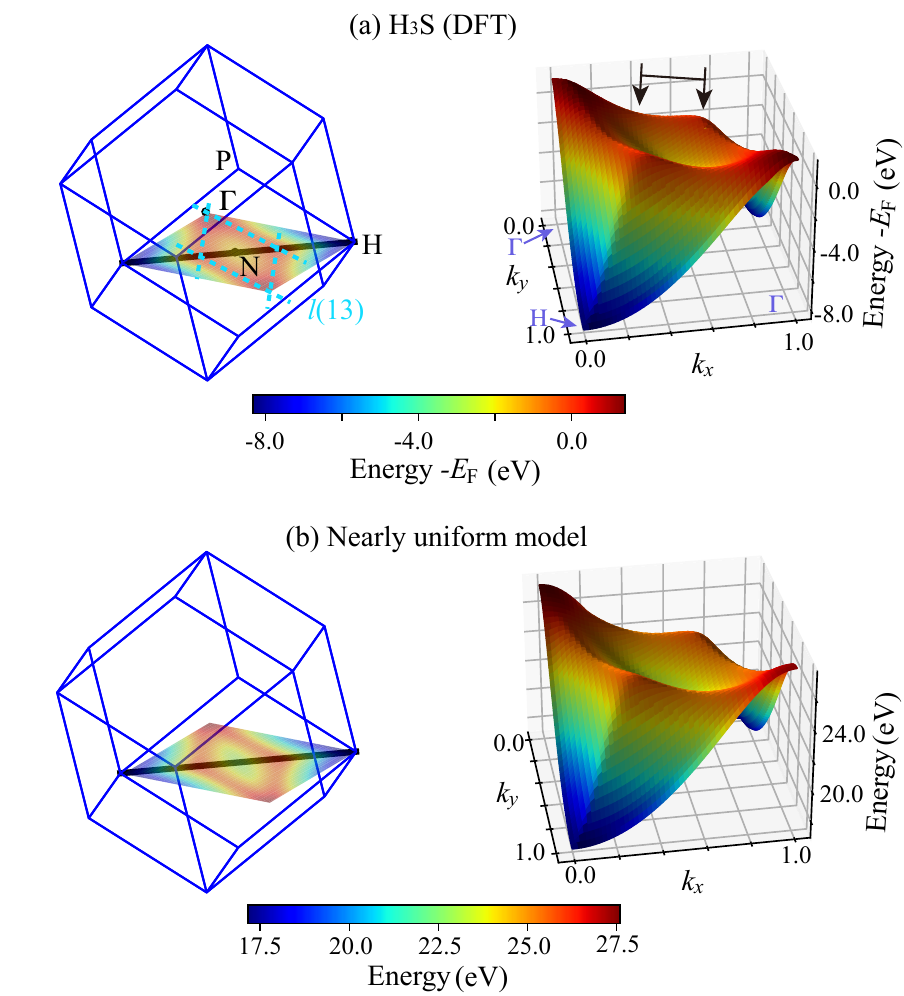}
   \caption{(a) Structure of the band responsible for the DOS peak in cubic H$_{3}$S in three dimensions, where its KS energy eigenvalues shown as heatmap on plane (0 0 1). Arrows in the right panel indicate the positions of the critical points that corresponds to the shoulders of the DOS peak\cite{Quan-Pickett-vHs-PRB2016}. (b) Same plot for the nearly uniform model.} 
   \label{fig:band-heat}
   \end{center}
 \end{figure*}

By applying extreme pressure on solid hydrogen sulfide, there forms a body-centered cubic (BCC) phase in space group $Im\bar{3}m$ ({\it No.} 229) with unconventional composition H$_{3}$S [Fig.~\ref{fig:bands_NF_BCC}(a)]. Superconductivity at 200 K has been discovered under pressures around 200~GPa~\cite{Eremets,Duan2014,Shimizu}. First-principles calculation has revealed that an anomalously narrow peak of the density of states (DOS) is located at the Fermi level, which is thought to be responsible for the strong superconducting pairing interaction~\cite{Duan2014,Errea-PRL2015,Bernstein-Mazin-PRB2015,Bianconi-EPL,Bianconi-NSM,Bianconi-scirep,Papaconstantopoulos-PRB2015, Akashi-PRB2015,Flores-Livas2016,Akashi-Magneli-PRL2016, Quan-Pickett-vHs-PRB2016}. The origin of the peak has been, however, enigmatic. Obvious flat bands like those in the nearest-neighbor tight-binding model~\cite{Jelitto1969,Souza-Marsiglio-simple-TB-PRB2016,Souza-Marsiglio-simple-TB2-IJMPB2017} have not been found in the first-principles band structures. The effective tight-binding models have been proposed~\cite{Bernstein-Mazin-PRB2015,Ortenzi-TBmodel-PRB2016,Quan-Pickett-vHs-PRB2016,Akashi2020} using the Slater-Koster~\cite{Slater-Koster} and first-principles Wannierization methods~\cite{MLWF1, MLWF2, wannier90}, with which some features in the band structure are successfully reproduced. But why the peak emerges from the models were not thoroughly investigated. The author has attempted this in a preceding paper~\cite{Akashi2020}. There he located saddle points extended in a loop form, termed saddle loop, and clarified that it is responsible for the DOS peak. However, to explain why the saddle loops emerge from the effective Wannier model, an intricate scenario based on farther neighbor hopping was necessary. In this section, we reexamine this issue with the knowledge of the intersections of Bragg planes. 

\begin{figure}[t!]
  \begin{center}
   \includegraphics[scale=0.22]{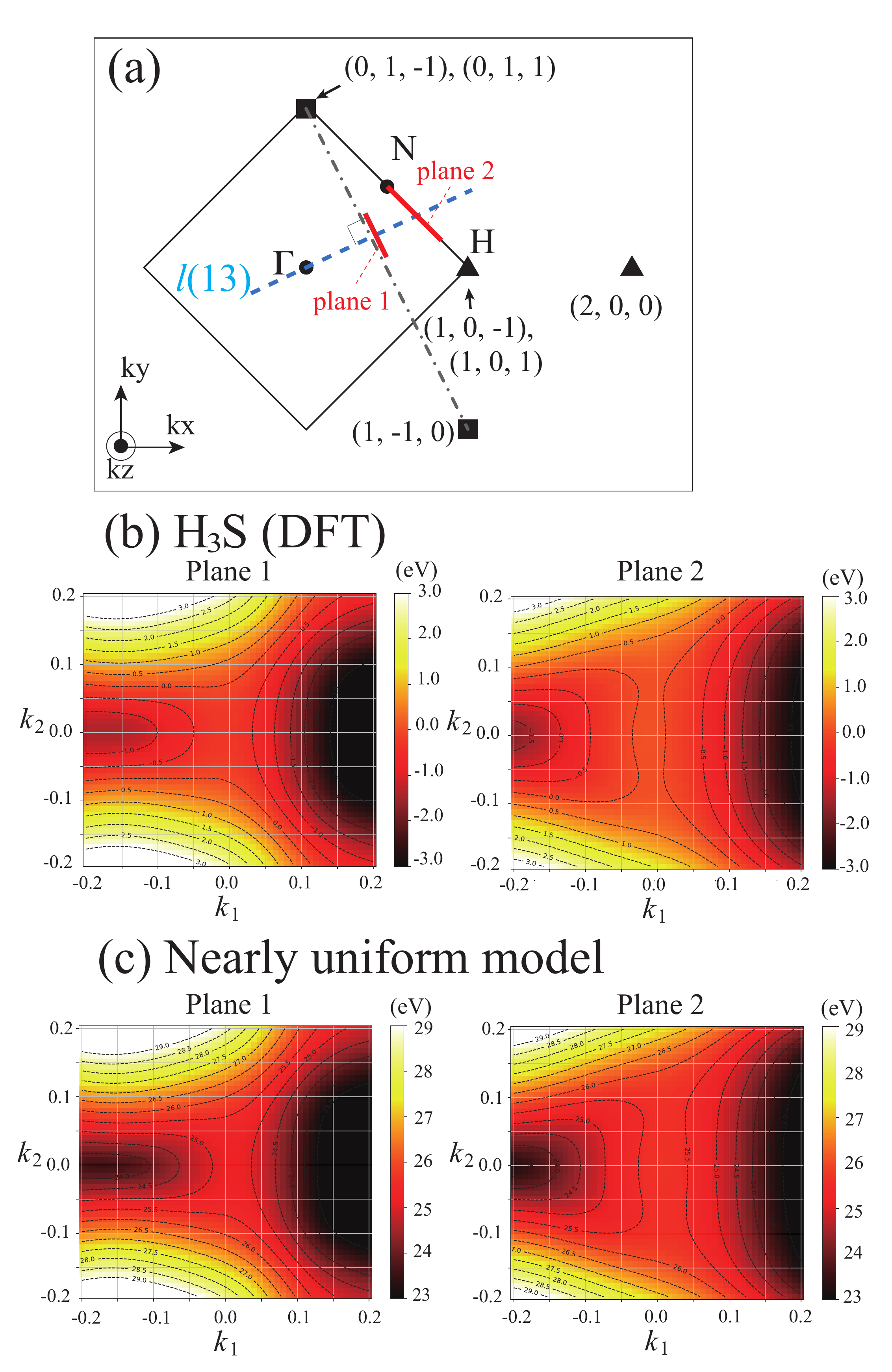}
   \caption{(a) Positions of the planes for the band calculations that cut the intersection $l(13)$. Square and triangle indicate the plane-wave branches that forms intersection $l(13)$ and those that significantly hybridize with the former, respectively. (b)(c) The heatmap plot of the band eigenvalues on planes 1 and 2.} 
   \label{fig:band-heat-planes}
   \end{center}
 \end{figure}

We start from observing a remarkable resemblance of the first-principles band structure of H$_{3}$S and empty lattice model one, shown in Fig.\ref{fig:bands_NF_BCC}(b)(c). The first-principles band structure was calculated using {\sc Quantum Espresso} code package~\cite{QE}, with the detailed condition being the same as that in Ref.~\cite{Akashi2020}. In addition to their apparent similarity, we find that the (near)-degenerate bands are assigned to the empty-lattice counterparts with perfect correspondence in the degrees of degeneracy [numbers in Fig.~\ref{fig:bands_NF_BCC}(b)]. This implies that the H$_{3}$S band structure could be well understood as the empty-lattice model with weak perturbations. In particular, the states contributing to the DOS peak in panel (c) seems to have emerged from the intersection line $l(13)$, that remains both in the SC and BCC lattices [Table~\ref{tab:paths}], and whose cross sections in the empty lattice limit are highlighted as square boxes in panel (b).

We further examine the detailed band structure of H$_{3}$S. We show in Fig.~\ref{fig:band-heat}(a) the Kohn-Sham eigenvalues of the 5th (in the ascending order) valence band that is responsible for the DOS peak. The eigenvalues are represented as a heat map on the plane $k_{z}=0$, from which it is clear that the extremum indicated in Fig.\ref{fig:bands_NF_BCC} (c) continuously extends inward the first Brillouin zone, with small band dispersion. Since this band has been found to be convex in the $k_{z}$ direction~\cite{Akashi2020}, this feature is classified as saddle loop. This saddle loop has been recently reproduced independently~\cite{Maarten-communication}. Note that the major portion of this looped structure has been first pointed out in a different look based on the conventional simple-cubic Brillouin zone~\cite{Akashi2020}. The edges of the loops match the intersection lines $l(13)$, which further support the hypothesis that the extended saddle originates from the intersection lines.


To the basic considerations in the previous section, electronic band dispersions around the extended singularities originating from the intersections should significantly depart from the quadratic forms. To confirm this we also show the band structures in two planar regions crossing $l(13)$ [Fig.~\ref{fig:band-heat-planes}(b)]. The contour plots clearly show high-order angular structures beyond quadratic. The characterization of the singularities based on the effective masses~\cite{Quan-Pickett-vHs-PRB2016} is hence incomplete for this system.

\begin{figure}[t!]
  \begin{center}
   \includegraphics[scale=0.45]{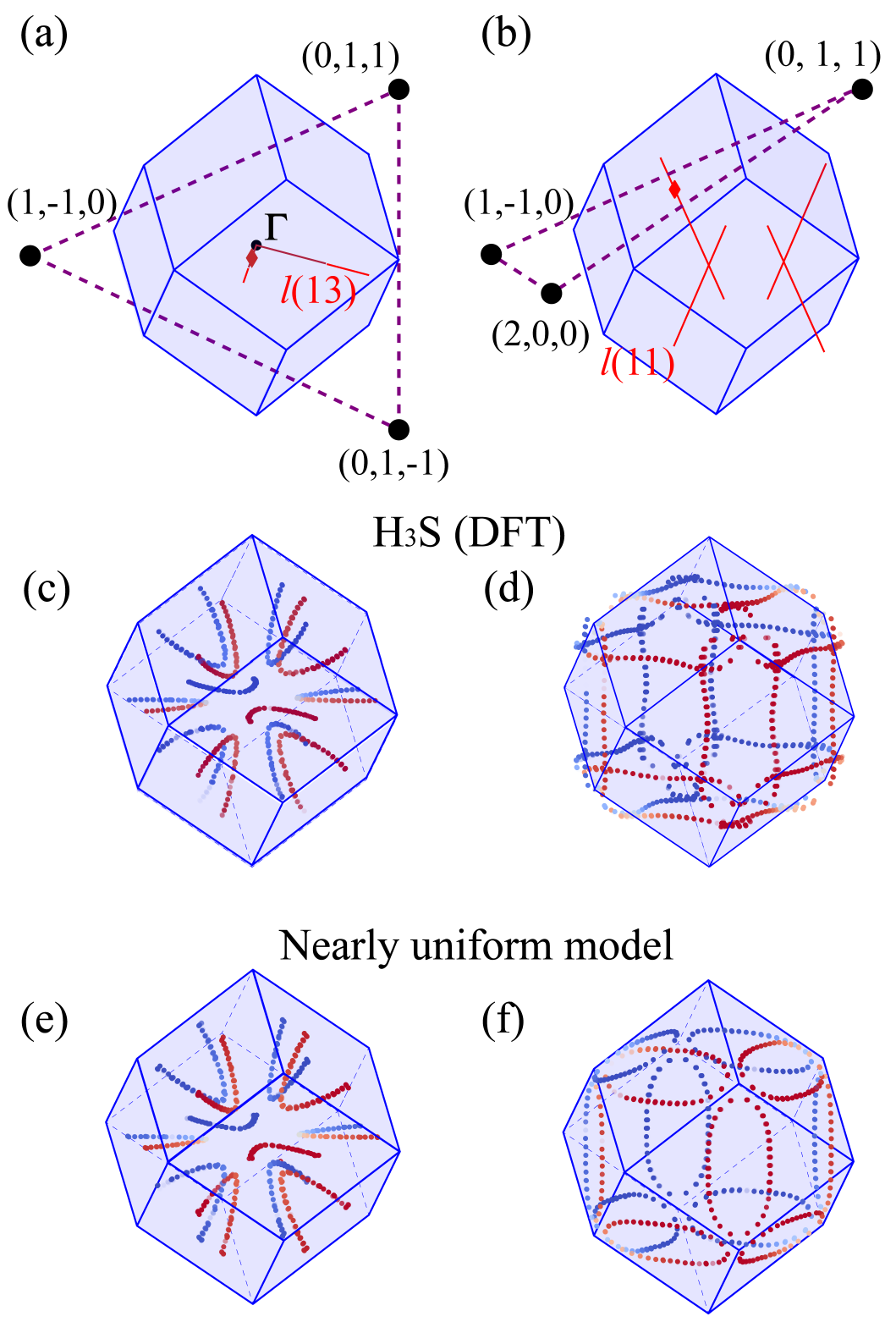}
   \caption{Positions of the Bragg intersections and Dirac nodal lines in H$_{3}$S. (a) [(b)] Intersection $l(13)$ [$l(11)$], with the corresponding $\Delta({\bf G}_{1},{\bf G}_{2},{\bf G}_{3})$ (black points) and ${\bf R}({\bf G}_{1},{\bf G}_{2},{\bf G}_{3})$ (red diamond). A symmetrically equivalent lines are also shown for comparison. (c) [(d)] Dirac nodal lines stemming from $l(13)$ [$l(11)$] calculated from first principles, colored in red (blue) for those in the front (back) half. (e) [(f)] Dirac nodal lines stemming from $l(13)$ [$l(11)$] with the nearly uniform model.} 
   \label{fig:DNL-sum}
   \end{center}
 \end{figure}

A remarkable consequence of the general theory is that the Bragg intersections may host the Dirac nodal lines, which has not been reported in the cubic H$_{3}$S to our knowledge. Executing a thorough search of the degeneracies, we confirmed this by finding undiscovered Dirac nodal lines formed among unoccupied bands that originate from the intersections $l(11)$ and $l(13)$, which are depicted in Fig.~\ref{fig:DNL-sum}. Intersection $l(13)$, the origin of the saddle loops, also induce the nodal loops that run on the plane $k_{z}=0$ almost in parallel [panels (a)(c)]. Those nodal loops appear in the standard band structure plot as points [dashed oval in Fig.~\ref{fig:bands_NF_BCC}~(c)]. Intersection $l(11)$, running on the Brillouin zone faces [Fig.~\ref{fig:DNL-sum}~(b)], induce fourfold degeneracy appearing as point in the P--H path marked by diamonds in Fig.~\ref{fig:bands_NF_BCC}~(b). A set of nodal lines in H$_{3}$S stemming from this [Fig.~\ref{fig:DNL-sum}~(d)] appears as a point in the middle of the P--H path [dotted oval in Fig.~\ref{fig:bands_NF_BCC}~(c)] and is located on the Brillouin-zone faces. These, say, $l(11)$ nodal lines cross with each other and the $l(13)$ nodal lines in the middle of H--N [\ref{fig:DNL-sum} (c)] at the point marked by dashed oval in Fig.~\ref{fig:bands_NF_BCC}~(c). The $l(11)$ nodal lines also cross with nodal lines running exactly on the Brillouin-zone edges, which are not shown here because they are supposed to originate from relatively trivial intersections on the edges. The mirror symmetries of the system confine the plotted nodal lines onto the $k_{z}=0$, Brillouin-zone faces and their symmetrical equivalents, all of which are invariant against their respective mirrors. The predicted existence of the Dirac nodal lines confirm the relevance of the Bragg intersection concept in H$_{3}$S.

Finally, we attempt to reproduce the band features found above using the nearly uniform electron model [Eq.~(\ref{eq:nearly-uniform})] on the BCC lattice. Through an only preliminary search, we found a set of potential components $\{\tilde{V}(|{\bf G}|\!\!=\!\!\sqrt{2}),\tilde{V}(|{\bf G}|\!\!=\!\! 2), \tilde{V}(|{\bf G}|\!\!=\!\!\sqrt{6})\}=(2\pi/a)^2\{-0.02, -0.05, 0.06\}$ that reproduced the crossing and anticrossing features with surprising accuracy as shown in Fig.~\ref{fig:bands_NF_BCC}(d). We were also successful in reproducing the abovementioned planar dependences of the first-principles extended saddles by the model [Figs.~\ref{fig:band-heat}(b) and \ref{fig:band-heat-planes}(c)], as well as the peaked DOS [Fig.~\ref{fig:bands_NF_BCC}(d)]. To our close analysis the band features concerned are formed by six plane-wave branches per intersection as depicted in Fig.~\ref{fig:band-heat-planes}(a). The companion nodal loops are also reproduced as well [Fig.~\ref{fig:DNL-sum}(e)]

We also list the remaining disagreements: (a) Subtle misordering of the degenerate bands was found at $\Gamma$. This ordering is irrelevant to the DOS peaking. In Ref.~\cite{Akashi2020} it was shown that the peaked DOS is most contributed by the k-point regions well away from $\Gamma$. This region corresponds to a portion of the ``ridge" of the extended saddle through $\Gamma$ to the BZ face, between the local minimum in the middle and local maximum at the face (arrows in [Fig.~\ref{fig:band-heat}(a)]). This tiny region dominates the DOS peak and therefore the $\Gamma$-point states have only a minor role in this sense. Also, (b) the model nodal loops induced by $l(11)$ were smaller than in H$_{3}$S and their crossing was not reproduced [Fig.~\ref{fig:DNL-sum}(f)]. This is due to the disappearance of the band crossing in P--H [see Figs.~\ref{fig:bands_NF_BCC}(c) and (d)]. Those disagreements could be remedied by futher fine tuning of the model, though we leave this out of the current scope.

Thus, we have seen that the major features of the band contributing to the DOS peak is interpretable from the nearly uniform electron model perspective. However, there is still missing how to reconcile it with the molecular orbital perspective~\cite{Heil-Boeri2015,Flores-Livas-review2020,Errea-NComm2021} that usually applies to molecular crystals. This point should be addressed in later studies. At this point, we stop by recalling a discussion by Kohn, on a Wannierization of nearly free-electron like systems~\cite{Kohn-Wannier}: Bloch-Wannier transformation to the plane-wave states yield function of form $\sim e^{i{\bf k}\cdot {\bf r}}/r$. The previously published Wannier functions in H$_{3}$S show remarkable sign changes and long tails~\cite{Quan-Pickett-vHs-PRB2016,Akashi2020}, which may indicate significant planewave-like characters of the relevant electronic states. 


\section{Summary}
In this paper, we have theorized effects of the intersections of the Bragg planes on the electronic single particle band structure. The translational symmetry determines the linearly extended regions in the k space in whose vicinity three or more plane-wave replica branches sensitively interact. Spatially extended higher order band anomalies are ensured to emerge around those regions, though their detailed features may depend on the Fourier components of the ionic potential. The present theory captures a mechanism of forming Dirac nodal lines and extended band critical points in three dimensional systems, which had long been hidden behind the difficulty of understanding the band structure, being a three dimensional hypersurface in four ($k_{x}-k_{y}-k_{z}-E({\bf k})$) dimensions. 

The Dirac nodal lines in elemental systems has been reported and discussed in relatively distinct literatures, especially after the rise of the topological nodal line trend~\cite{Li2016,Hirayama2017,Allen-Pickett2018}. Such studies wholly focused on the specific properties of the discovered nodal lines like the topological indices and surface states. The Bragg intersection concept obviously serves a unified explanation of a considerable part of such nodal lines as to why and how they emerge {\it there} in the Brillouin zones, as well as possible companion van Hove singularities.


The new theoretical view have helped an unprecedented complete characterization of the first-principles electronic band structure near the Fermi level in cubic H$_{3}$S. The current analysis encourages further studies of possible anomalous low-energy phenomena from the non-quadratic band dispersion~\cite{Yuan-Fu-vHS, Markiewicz-review} and complementary understanding of this system from nearly uniform electron perspective. Re-investigation on the electron-phonon perturbation theories with precise treatment of the revealed band features are also of interest~\cite{Durajski-PhysicaC2015,Sano-vHS-PRB2016,Capitani-Nphys2017,Ghosh-H3S-ph-break}.

\section*{Acknowledgment}
This work was supported by JSPS KAKENHI Grant Numbers 20K20895 and 23K03313 from Japan Society for the Promotion of Science (JSPS). The author thanks to Yu-ichiro Matsushita, Taichi Kosugi, Yusuke Nishiya and Hung Ba Tran for useful comments. Some of the calculations were performed at the Supercomputer Center in ISSP, the University of Tokyo.



\bibliography{reference}

\end{document}